\documentclass[%
 aip,
 amsmath,amssymb,nofootinbib,
 reprint,%
 scrartcl,
]{revtex4-1}

\usepackage{xcolor}
\usepackage{graphicx}
\usepackage{dcolumn}
\usepackage{bm}
\usepackage[utf8]{inputenc}
\usepackage[T1]{fontenc}
\usepackage{mathptmx}
\usepackage{amsmath}
\usepackage{mathtools}
\usepackage[none]{hyphenat}
\DeclareUnicodeCharacter{221A}{-}
\DeclareUnicodeCharacter{00B0}{-}

\usepackage{etoolbox}

\renewcommand{\thefootnote}{\roman{footnote}}

\begin{document}
\preprint{AIP/123-QED}

\title{Anti-reflection Coated Vacuum Window for \\ the Primordial Inflation Polarization ExploreR (PIPER) balloon-borne instrument}

\author{Rahul~Datta}
\affiliation{%
Department of Physics and Astronomy, Johns Hopkins University, 3701 San Martin Drive, Baltimore, MD 21218~USA
}%
\email{rdatta2@jhu.edu} 

\author{David~T.~Chuss}
\affiliation{%
Department of Physics, Villanova University, 800 Lancaster Avenue, Villanova, PA 19085~USA
}%
\author{Joseph~Eimer}
\affiliation{%
Department of Physics and Astronomy, Johns Hopkins University, 3701 San Martin Drive, Baltimore, MD 21218~USA
}%
\author{Thomas~Essinger-Hileman}
\affiliation{%
NASA Goddard Space Flight Center, 8800 Greenbelt Road, Greenbelt, MD 20771~USA
}%
\author{Natalie~N.~Gandilo}
\affiliation{%
NASA Goddard Space Flight Center, 8800 Greenbelt Road, Greenbelt, MD 20771~USA
}%
\author{Kyle~Helson}
\affiliation{%
NASA Goddard Space Flight Center, 8800 Greenbelt Road, Greenbelt, MD 20771~USA
}%
\affiliation{%
The Center for Space Sciences and Technology, The University of Maryland Baltimore County, 1000 Hilltop Cir, Baltimore, MD 21250
}%
\author{Alan~J.~Kogut}
\affiliation{%
NASA Goddard Space Flight Center, 8800 Greenbelt Road, Greenbelt, MD 20771~USA
}%
\author{Luke~Lowe}
\affiliation{%
NASA Goddard Space Flight Center, 8800 Greenbelt Road, Greenbelt, MD 20771~USA
}%
\author{Paul~Mirel}
\affiliation{%
NASA Goddard Space Flight Center, 8800 Greenbelt Road, Greenbelt, MD 20771~USA
}%
\author{Karwan~Rostem}
\affiliation{%
NASA Goddard Space Flight Center, 8800 Greenbelt Road, Greenbelt, MD 20771~USA
}%
\author{Marco~Sagliocca}
\affiliation{%
NASA Goddard Space Flight Center, 8800 Greenbelt Road, Greenbelt, MD 20771~USA
}%
\author{Danielle~Sponseller}
\affiliation{%
Department of Physics and Astronomy, Johns Hopkins University, 3701 San Martin Drive, Baltimore, MD 21218~USA
}%
\author{Eric~R.~Switzer}
\affiliation{%
NASA Goddard Space Flight Center, 8800 Greenbelt Road, Greenbelt, MD 20771~USA
}%
\author{Peter~A.~Taraschi}
\affiliation{%
NASA Goddard Space Flight Center, 8800 Greenbelt Road, Greenbelt, MD 20771~USA
}%
\author{Edward~J.~Wollack}
\affiliation{%
NASA Goddard Space Flight Center, 8800 Greenbelt Road, Greenbelt, MD 20771~USA
}%

\date{\today}

\begin{abstract}
Measuring the faint polarization signal of the cosmic microwave background (CMB) not only requires high optical throughput and instrument sensitivity but also control over systematic effects. Polarimetric cameras or receivers used in this setting often employ dielectric vacuum windows, filters, or lenses to appropriately prepare light for detection by cooled sensor arrays. These elements in the optical chain are typically designed to minimize reflective losses and hence improve sensitivity while minimizing potential imaging artifacts such as glint and ghosting. The Primordial Inflation Polarization ExploreR (PIPER) is a balloon-borne instrument designed to measure the polarization of the CMB radiation at the largest angular scales and characterize astrophysical dust foregrounds. PIPER$^{\prime}$s twin telescopes and detector systems are submerged in an open-aperture liquid helium bucket dewar. A fused-silica window anti-reflection (AR) coated with polytetrafluoroethylene (PTFE) is installed on the vacuum cryostat that houses the cryogenic detector arrays. Light passes from the skyward portions of the telescope to the detector arrays though this window, which utilizes an indium seal to prevent superfluid helium leaks into the vacuum cryostat volume. The AR coating implemented reduces reflections from each interface to $<1\%$ compared to $\sim10\%$ from an uncoated window surface. The AR coating procedure and room temperature optical measurements of the window are presented. The indium vacuum sealing process is also described in detail and test results characterizing its integrity to superfluid helium leaks are provided.
\end{abstract}

\maketitle

\section{Introduction}
Vacuum windows are widely used in astronomical measurement systems in the radio through the terahertz wavebands that employ cryogenically cooled detectors and optics. Common application examples include use in radio astronomy receivers, airborne remote sensing systems, and suborbital instruments for characterization of the cosmic microwave background (CMB). These receiver systems require a physical barrier or window that is transparent over the desired waveband at the interface between the external ambient environment and the instrument's internal vacuum space. This window interface simultaneously holds vacuum against ambient pressure and enables the desired optical observations.

The window's geometry, service environment, and optical response are intimately tied to the material properties used in the design. Over the observing band, minimal in-band dielectric absorption and low total emissivity is preferred. An anti-reflection (AR) coating is often employed to mitigate the loss of sensitivity and image artifacts arising from reflections. In practice, there is a trade between the magnitude of the dielectric function of the window material and the number of layers or complexity of the AR coating structures required to provide high transmission over a desired bandwidth~\cite{2018JLTP..193..876C,2018SPIE10708E..43N}. In addition, the window needs to provide a reliable mechanical barrier to withstand the pressure differential at the vacuum--ambient interface and accommodate thermally induced stresses in the assembly.  Two basic implementation strategies, the use of transparent deformable membranes and rigid structural elements, are commonly encountered in practice. 
 
Prior deformable vacuum window designs have primarily relied upon the use of low index polymeric materials. Examples include high-density polyethylene (HDPE)~\cite{Sharp2008,Harper2018} and reinforced ultra-high-molecular-weight polyethylene (UHMWPE)~\cite{Barkats2018} films, which are well-suited for room temperature instrument window interfaces operating at microwave through sub-millimeter wavelengths. Composite window structures using low-density closed-cell foams to provide structural support for relatively thin plastic vacuum barriers can also be found in this setting~\cite{Kerr1992,Runyan2003}. While successfully deployed in the field, cold flow, gas permeability, and changes in material properties with temperature present potential challenges for reliable cryogenic application of this design strategy.
 
Structural vacuum window implementations tend to utilize higher index materials with low attenuation, high thermal conductivity, and high mechanical strength. Sapphire, alumina, diamond, and compensated silicon have been considered for cryogenically cooled windows in high-power millimeter-wave applications~\cite{parshin95}. In the microwave through sub-millimeter wavebands z-cut crystalline quartz windows have been widely used in astronomical instrumentation~\cite{Koller2001}. Similarly, commonly available industrial grade fused silica glass has also been used as a window material for ground based CMB applications~\cite{10.1117/12.459274}. It is notable that crystalline media are generally preferred over amorphous dielectric media in this spectral regime from a dielectric loss perspective at cryogenic temperatures~\cite{1991AdPhy..40..719G,Petzelt2003SubmillimetreAI,1982IJIMW...3..929A}. In this approach, the optical path loss is essentially dictated by the window thickness required to prevent failure of the window structure~\cite{Ventsel2001}. For cryogenic applications, differences in the coefficient of thermal expansion between the window and the AR coating can present an implementation concern. If unaddressed, the resultant thermally induced stress can potentially lead to delamination of the AR coating or compromise the barrier's structural integrity.

The specific window implementation presented here is intended for use on the Primordial Inflation Polarization ExploreR (PIPER) balloon-borne instrument~\cite{10.1117/12.2313874,10.1117/12.2231109}. This system is designed to map the polarization of the CMB on the largest angular scales and characterize dust foregrounds by observing a large fraction of the sky in four frequency bands in the range 200--600~GHz~\cite{Kogut2020Primordial}. The PIPER receiver~\cite{2019RScI...90i5104S,Switzer2020Primordial} is submerged in liquid helium, which becomes a superfluid during flight due to the low ambient pressure at balloon altitudes, and represents additional challenges for realization of the vacuum window assembly. The window and all of its interfaces need to be impervious to superfluid helium and able to reliably survive multiple thermal cycles. 

This work focuses on the window and vacuum seal implementation for PIPER's 200~GHz observing band. The paper is organized as follows: in Section~II, design considerations for the proposed system application are discussed. The optical model used for evaluation of the window and AR-coating response is described in Section~III. In Section~IV, experimental characterization of the optical materials in use and the final AR-coated window assembly is presented. The window installation employing a vacuum seal robust to superfluid helium leaks is described in Section~V. The setup for thermal cycling tests and characterization of the vacuum window leakage rates are presented in Section~VI.  A summary can be found in Section~VII. 

\begin{figure}
\centering\includegraphics[width=1.0\linewidth,keepaspectratio]{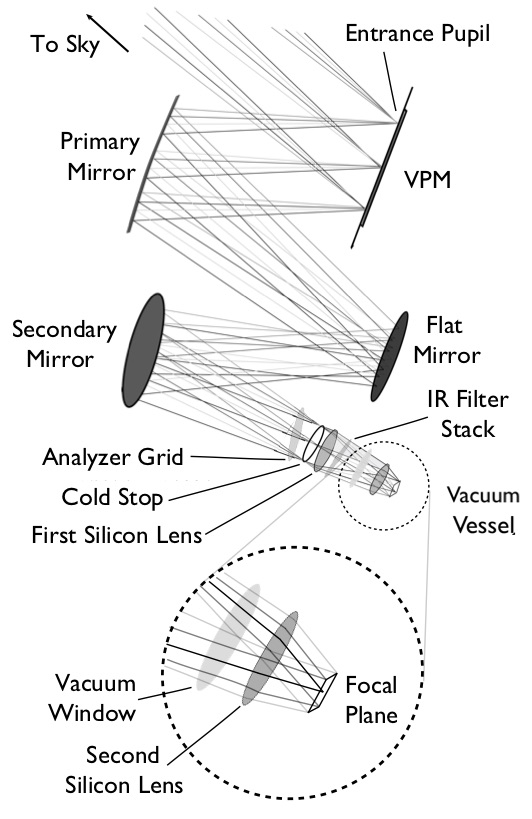}
\caption{{A simplified ray trace of the PIPER optics. PIPER employs twin telescopes for simultaneously measuring Stokes Q and U, only one of the two independent optical paths is shown here. PIPER’s reflective fore-optics are in an off-axis Gregorian-like configuration~\cite{Eimer2010} with a Variable-delay Polarization Modulator (VPM)~\cite{2014RScI...85f4501C} as the first optical element, primary mirror, folding flat mirror, and secondary mirror. The light is focused onto the detectors by silicon re-imaging optics comprising two anti-reflection coated silicon lenses~\cite{2018JLTP..193..876C}. The superfluid-tight vacuum vessel houses the receiver, which contains the second silicon lens, sub-Kelvin cooled detector package, cold readout components, and the cooling system~\cite{2019RScI...90i5104S}. The vacuum vessel houses both focal planes, with separate  windows for each  path.} }
\end{figure}

\section{Window Design Considerations}
\vskip1sp
A ray trace schematic of the PIPER optics is shown in Fig.~1 highlighting the main optical components~\cite{Eimer2010}. Light is focused onto the detectors by two silicon lenses, one in front of the vacuum window and the other just after the window within the superfluid helium tight vacuum vessel that contains the focal plane arrays~\cite{2019RScI...90i5104S}. The window is located at an intermediate focus and illuminated by incidence angles ranging from 0 to 16$^{\circ}$ with the optical axis normal to its surface.

\begin{figure*}
\centering\includegraphics[width=0.9\linewidth,keepaspectratio]{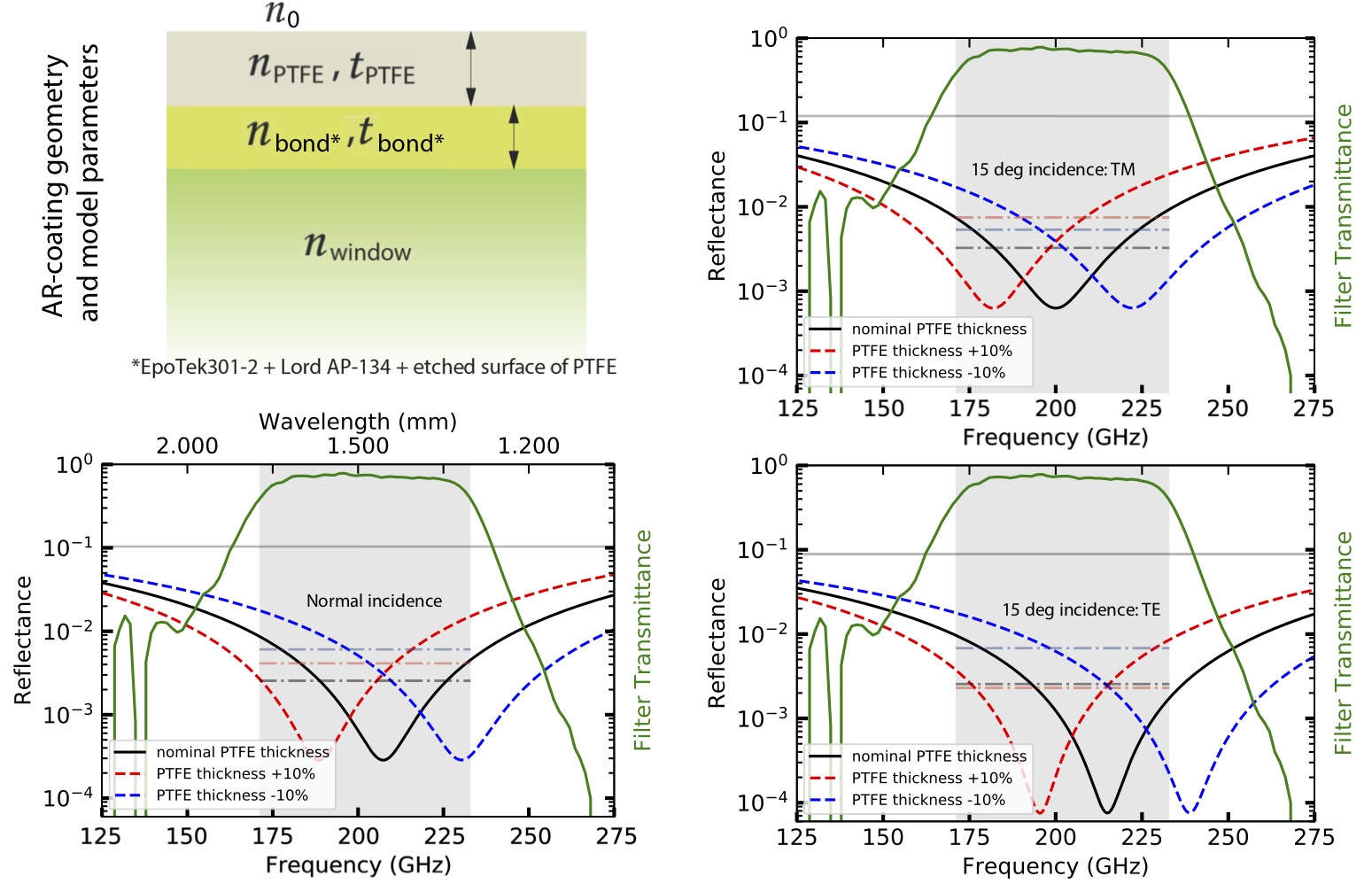}
\caption{{\it{Top Left}}: Schematic representation of the AR-coating layers on a dielectric half-space used to model the reflectance of the window. For simplicity lossless materials are adopted in this simulation. {\it{Bottom Left}}: Analytical reflectance at normal incidence (black) from a window AR-coated with a 254~$\mu$m thick layer of PTFE. The simulation assumes a 12.7~$\mu$m thick bond layer. For reference, the filter-stack transmittance for the PIPER instrument's $200$\,GHz channel is indicated in green. The grey shaded region represents the filter's half-power bandwidth, which spans 171--233~GHz. The design tolerances to $\pm$10$\%$ deviation in the PTFE layer thickness are also shown in red and blue, respectively. The horizontal lines represent the reflectance averaged over the band of interest shown in grey. {\it{Right}}: Analytical reflectance at 15 degree oblique incidence for Transverse Magnetic (TM, top panel) and Transverse Electric (TE, bottom panel) modes of polarized light for the nominal PTFE layer thickness as well as $\pm$10$\%$ deviation from nominal. The horizontal grey lines in all three plots correspond to the reflectance in the absence of an AR coating.}
\end{figure*}

Noting that the materials commonly used in deformable membranes are not impervious to superfluid helium, a rigid structural element is adopted for the window design. Crystalline quartz and fused silica were the primary candidates considered due to their optical and mechanical properties in the bulk form. Their modest index of refraction allows implementation of an AR coating operating over $\sim$30\% fractional bandwidth with a single layer realized from a readily available polymer such as PTFE~\cite{Koller2001}. A survey of the dielectric properties of amorphous silica and crystalline quartz and a discussion of their microwave through infrared absorption can be found in Cataldo {\it et al.}~\cite{Cataldo:16}. 

Industrial grade fused silica glass substrate\footnote{www.mcmaster.com; fused quartz window 7.500$^{\prime\prime}$ diameter, 0.375$^{\prime\prime}$ thick (part number 1357T999)} was selected over crystalline quartz as the transparent structural material for the PIPER vacuum windows due to its availability in the aperture size of interest. The diameter of the window clear aperture is 19.1~cm driven by the size of the optical beam, which spans 15.2~cm in diameter at the location of the window. The thickness of the fused silica window is approximately 9.5~mm, chosen so as to support a pressure of an atmosphere with a margin of safety of three. The coated section of the window should span the beam which sets the lower limit on the coating diameter of 15.2~cm. In order to ensure a proper vacuum seal, the edge of the window should be clear and smooth, which sets an upper limit on the coating diameter of approximately 17.2~cm. A nominal diameter of 16~cm for the AR-coated region of the window was targeted in fabrication.

\section{Window and AR Coating Optical Model}
\vskip1sp
In designing a layered AR structure, the index of refraction and thickness of the coating layers are chosen such that the reflection amplitudes arising from dielectric interfaces destructively interfere. For a lossless single-layer AR coating between vacuum and a substrate with refractive index $n$, perfect cancellation at normal incidence occurs when the layer's refractive index is $\sqrt{n}$ and thickness provides a quarter-wavelength optical delay. The refractive index of PTFE~\cite{Goldsmith1998} is $\approx$1.42, close to $\sqrt{n}$ for fused silica, and a suitable choice as a single-layer AR coating.

\begin{figure}
\centering\includegraphics[width=1\linewidth,keepaspectratio]{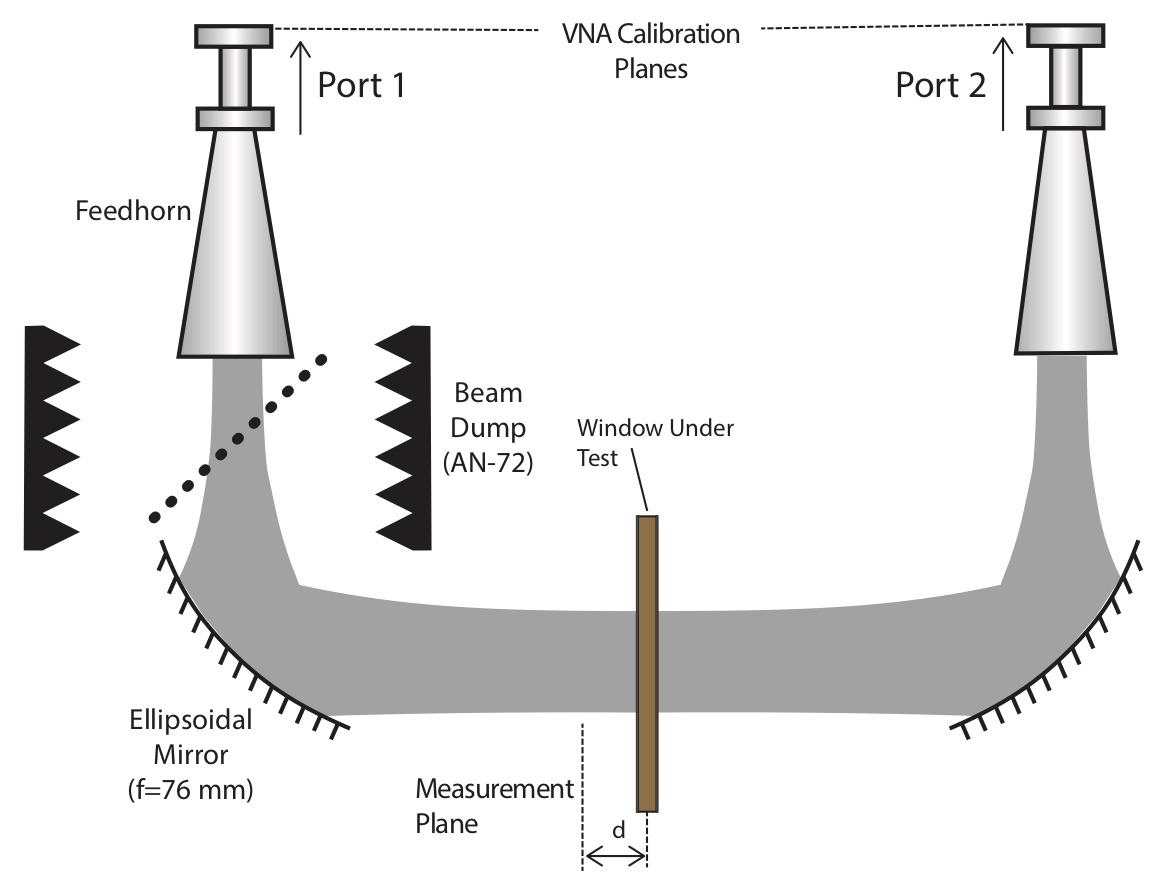}
\caption{(Adapted with permission from D. T. Chuss, K. Rostem, E. J. Wollack, L. Berman, F. Colazo, M. DeGeorge,K. Helson, and M. Sagliocca, Review  of  Scientific  Instruments, Vol.~88, Page~104501, (2017),~\cite{2017RScI...88j4501C}.) Reflection and transmission measurement setup using a Vector Network Analyzer (VNA). The sample-under-test was placed approximately at the focus between the ellipsoidal mirrors. }

\end{figure}

In practice, an analytical model is used for evaluation of the window's optical response and to explore the sensitivity of the nominal design parameters to fabrication tolerances. The reflectance is computed for a plane-parallel stack of $m$ dielectric layers of physical thickness $t_\mathrm{m}$ illuminated by a collimated beam. The net reflectance for normal incidence at the $j$-th interface is given by the recursion relation~\cite{Orfanidis}: 
\begin{eqnarray}
&&\Gamma_\mathrm{j} = \frac{\rho_\mathrm{j}+\Gamma_\mathrm{j+1}e^{-2ik_\mathrm{j}t_\mathrm{j}}}{1+\rho_\mathrm{j}\Gamma_\mathrm{j+1}e^{-2ik_\mathrm{j}t_\mathrm{j}}}, \ \ \ \mathrm{j} = m, m-1,....,1\nonumber\\
 &&\textnormal{where,}\ \ \rho_\mathrm{j} = \frac{\hat{n}_\mathrm{j-1}-\hat{n}_\mathrm{j}}{\hat{n}_\mathrm{j-1}+\hat{n}_\mathrm{j}},\ k_\mathrm{j} = \frac{2\pi\hat{n}_\mathrm{j}}{\lambda}, \ \ \ \hat{n}_\mathrm{0} = \hat{n}_\mathrm{m+1} = 1
\label{eq:ref}
\end{eqnarray}
and initialized by $\Gamma_\mathrm{m+1} = \rho_\mathrm{m+1}$. In Eq.~\ref{eq:ref} the complex refractive index, $\hat{n} \equiv n + i \kappa$ is related to the relative dielectric function, $\hat{\epsilon}_r \equiv \epsilon' + i \epsilon''$, by the following relation:
\begin{eqnarray}
  \hat{n}^{2} = (n^{2} - \kappa^{2}) +  i 2n \kappa =
  \hat{\epsilon}_{r}. 
\label{eq:nhat}
\end{eqnarray}
The dielectric permittivity, $\hat{\epsilon}_{r}$, is in general frequency dependent, however, over the relatively narrow spectral range considered here, $\epsilon'$ and $\epsilon''$ can be considered constants. The ratio of the imaginary to the real part of the relative dielectric function gives a measure of the dissipation in the medium, commonly referred to as the loss tangent of the material:
\begin{eqnarray}
\tan \delta_{e} = \epsilon^{\prime\prime}/\epsilon^{\prime} = 2n \kappa/(n^{2} - \kappa^{2}).
\end{eqnarray}
The model defined in Eq.~\ref{eq:ref} can be readily extended to accommodate oblique incidence~\cite{Orfanidis} and was used to investigate the polarization dependence of the dielectric stack. An alternative optical modeling approach to the one adopted in this work is the transmission line or ABCD matrix method~\cite{Yeh2005,Goldsmith1998,Byrnes2016MultilayerOC}.

In specifying the window optical response the goal was to minimize reflectance over the instrument frequency bandwidth while limiting the coating to a single commercially-available thickness of PTFE sheet ({\it{e.g.}}, discrete thicknesses of 2, 4, 5, 10, 15, etc. in thousandths of an inch) uniformly etched on one surface to aid adhesion to the window surface. The etching process\footnote{www.professionalplastics.com/TEFLONETCHED} involved surface treatment using a mixture of sodium and ammonia that is formulated to create a bondable surface. The bond is realized with Epo-Tek\footnote{Epoxy Technology, Inc. (www.epotek.com)}~301-2 epoxy~\cite{Koller2001} and adhesion promoter~\cite{Lau:06} Lord AP-134~\footnote{www.chemical-concepts.com/lord-ap-134.html}. This combination improves adhesion of the AR coating to the silica substrate and has been observed to enable the optical structure to survive multiple cryogenic cycles. The dielectric function of the Epo-Tek~301-2 epoxy in the far-IR, $\hat{\epsilon}_r \simeq 3.7 + 0.1i$, is reported in Munson {\it et al.}~\cite{Munson:17} and was adopted for modeling purposes. 

\begin{figure*}
\centering\includegraphics[width=0.95\linewidth,keepaspectratio]{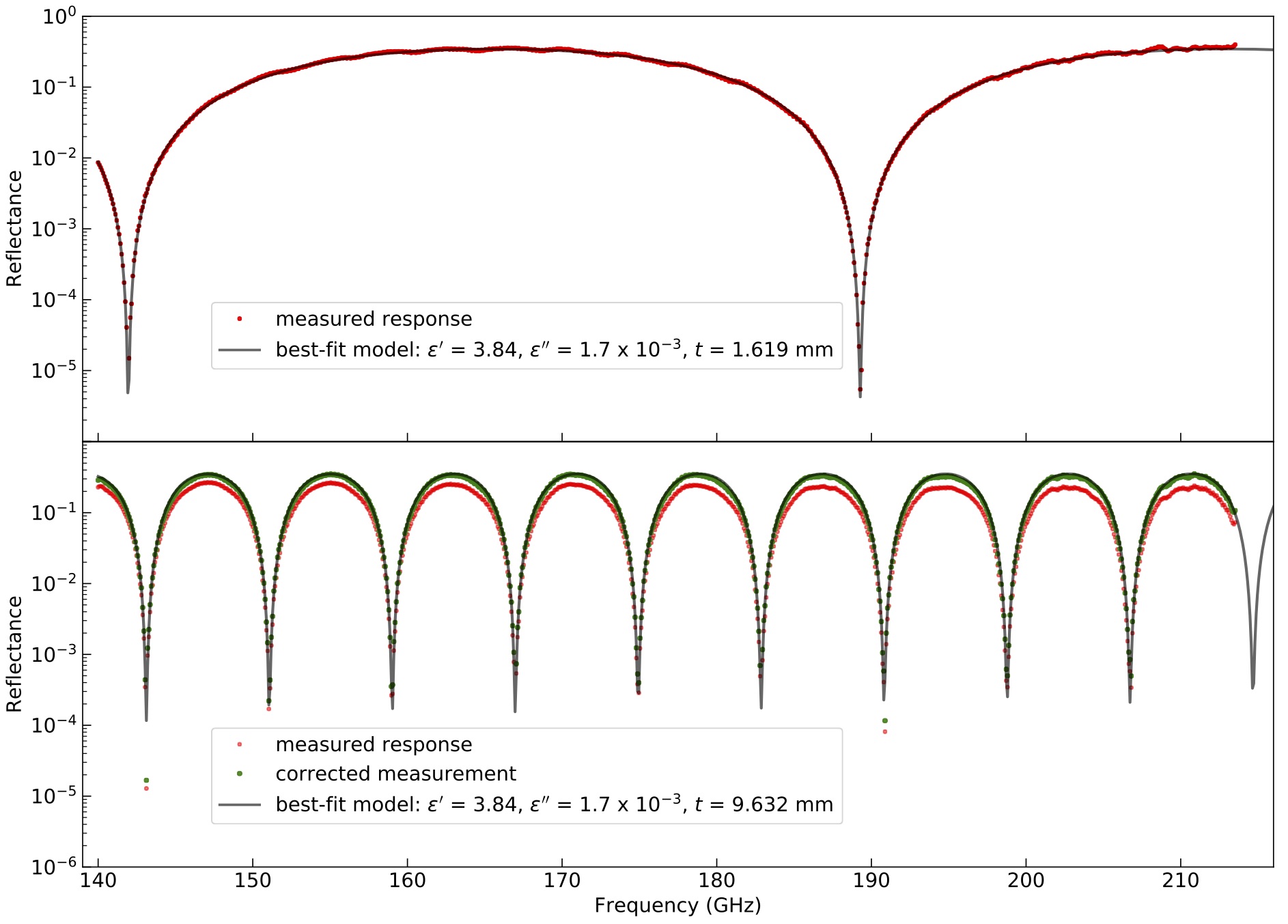} 
\caption{{\it{Top:}} Measured reflectance (red dots) from a 1.64~mm thick bulk fused silica sample along with the best-fit analytical model (grey line) with parameters $\epsilon'$, $\epsilon''$, and thickness $t$. The fit values are noted in the legend. {\it{Bottom:}} Measured reflectance (red dots) from a 9.69~mm thick sample and the corrected reflectance (green dots), which accounts for the difference between the modeled plane wave illumination of the sample and a focused beam~\cite{1017671} in the measurement setup. The correction follows the prescription outlined in Section IV.A. The analytical model (grey line) assumes $\epsilon'$ to be fixed, its value given by the best-fit value for the thin sample in the top panel. The numerical coefficients in the correction function, $\epsilon''$, and thickness $t$ are free parameters while fitting an analytical model scaled by the inverse of the correction function to the data. The correction function determined from the fit is then applied back to the raw measured data (red dots) to get the corrected measured reflectance (green dots). The same exercise is repeated with two additional samples. The complex dielectric function of the bulk silica as inferred from the fits to these measurements is $\hat{\epsilon}_r$ = (3.84$\pm$0.02) + (0.0021$\pm$0.0007)$i$, where the uncertainty in $\epsilon''$ is the error on the mean from the individual fits to the data from the four bulk samples measured. The reflectance spectra as well as the analytical model have identical sampling and are smoothed with an 11-point ({\it{i.e.}}, 0.9075 GHz) Savitsky-Golay filter~\cite{SavitzkyGolay:1964} for the purpose of presentation. } 
\end{figure*}

From the available physical thicknesses of PTFE, 0.010$^{\prime\prime}$ (254 $\mu$m) is very close to quarter-wavelength electrical delay at 200~GHz. Using this nominal value for the coating thickness the response of the window was modeled assuming notional parameters for the bond layer. The model was evaluated to investigate the influence of PTFE thickness variations and sensitivity to incident polarization. For simplicity, lossless dielectrics were adopted. See Fig.~2 (top left panel) for a schematic of the modeled window geometry and index of refraction for each layer. The bottom left panel of Fig.~2 shows the predicted reflectance at normal incidence assuming $n_{\mathrm{window}}$~=~1.95, $n_{\mathrm{PTFE}}$~=~1.42, $t_{\mathrm{PTFE}}$~=~254~$\mu$m, $n_{\mathrm{bond}}$~=~$\sqrt{3.7}$, and $t_{\mathrm{bond}}$~=~12.7~$\mu$m. In all the responses evaluated, the reflectance averaged over the 171--233~GHz half-power bandwidth was well below 1$\%$.

\section{Reflection and Transmission Measurements}
In this section, reflection measurements of bulk silica samples and experimental methods employed are described. Measurements of the complex dielectric function of the PTFE used for the coating are also presented.  The reflectance from a prototype silica vacuum window anti-reflection coated on both sides with a layer of PTFE is measured. The procedure for applying the coating is described in detail in Appendix A. In comparing these measurements of the AR-coated window structure to the modeled optical response, the derived material properties of bulk dielectric samples are used as constraints. All the measurements described were carried out at room temperature. Given the materials in use are amorphous, their dielectric response will change by a perturbation related to the change in density upon cooling to cryogenic temperatures~\cite{Lamb1996MiscellaneousDO}. 

\subsection{Dielectric Characterization: Bulk Silica } 
The reflectance and transmittance of bulk fused silica samples of various thicknesses were measured using an Agilent PNA-X vector network analyzer (VNA) coupled to the free space quasioptical setup~\cite{2017RScI...88j4501C} shown in Fig.~3.  A multi-tier calibration procedure was employed following Chuss {\it et al.}~\cite{2017RScI...88j4501C}. The VNA was calibrated at the waveguide ports using WR05 Thru-Reflect-Line (TRL) standards over a spectral range spanning 140--$220$\,GHz. The test system was aligned by adjusting the position and orientation of the mirrors to maximize the observed transmission prior to insertion of the sample. A translation stage with a digital micrometer readout was used to hold the sample and measure the reflectance over a discrete set of free space delays corresponding to 0, 0.2$\lambda_\mathrm{0}$, 0.4$\lambda_\mathrm{0}$, 0.6$\lambda_\mathrm{0}$, 0.8$\lambda_\mathrm{0}$, and $\lambda_\mathrm{0}$, where $\lambda_\mathrm{0} = 1670\,\mu{\rm m}$ is the free space wavelength at the center of the VNA spectral range. This calibration procedure enables the instrumental response of the free space VNA test configuration to be separated and removed from the sample-under-test.  

\begin{table*}
\caption{Best-fit model parameters for the measured bulk fused silica samples. The measurements are averaged readings obtained with a micrometer over six locations on the sample. The uncertainties are standard errors on the fit parameters. For sample 1 the magnitude $\epsilon^{\prime}$ is derived from fit and then adopted as a parameter for the other samples.}
\begin{center}
\begin{tabular}{lclclclclclclcl}
\hline
&&& sample 1 &&& sample 2 &&& sample 3 &&& sample 4\\ 
\hline
measured $t$ (mm) &&& 1.64$\pm$0.01 &&& 6.16$\pm$0.02 &&& 9.69$\pm$0.03 &&& 12.73$\pm$0.03 \\
fit $t$ (mm) &&& 1.619$\pm$0.002 &&& 6.130$\pm$0.01 &&& 9.632$\pm$0.02 &&& 12.753$\pm$0.03 \\
$\epsilon^{\prime}$ &&& 3.84$\pm$0.01 &&& (fixed) &&& (fixed) &&& (fixed) \\
$\epsilon^{\prime\prime} \times10^{3}$ &&& 1.7$\pm$0.1 &&& 1.7$\pm$0.1 &&& 1.7$\pm$0.1 &&& 3.1$\pm$0.1 \\
$\tan\delta_{e} \times10^{4}$ &&& 4.4$\pm$0.2 &&& 4.4$\pm$0.2 &&& 4.5$\pm$0.2 &&& 8.3$\pm$0.3 \\
\hline                 
\end{tabular}\\
\end{center}
\end{table*}
\label{tab:model}

\begin{figure}
\centering\includegraphics[width=0.95\linewidth,keepaspectratio]{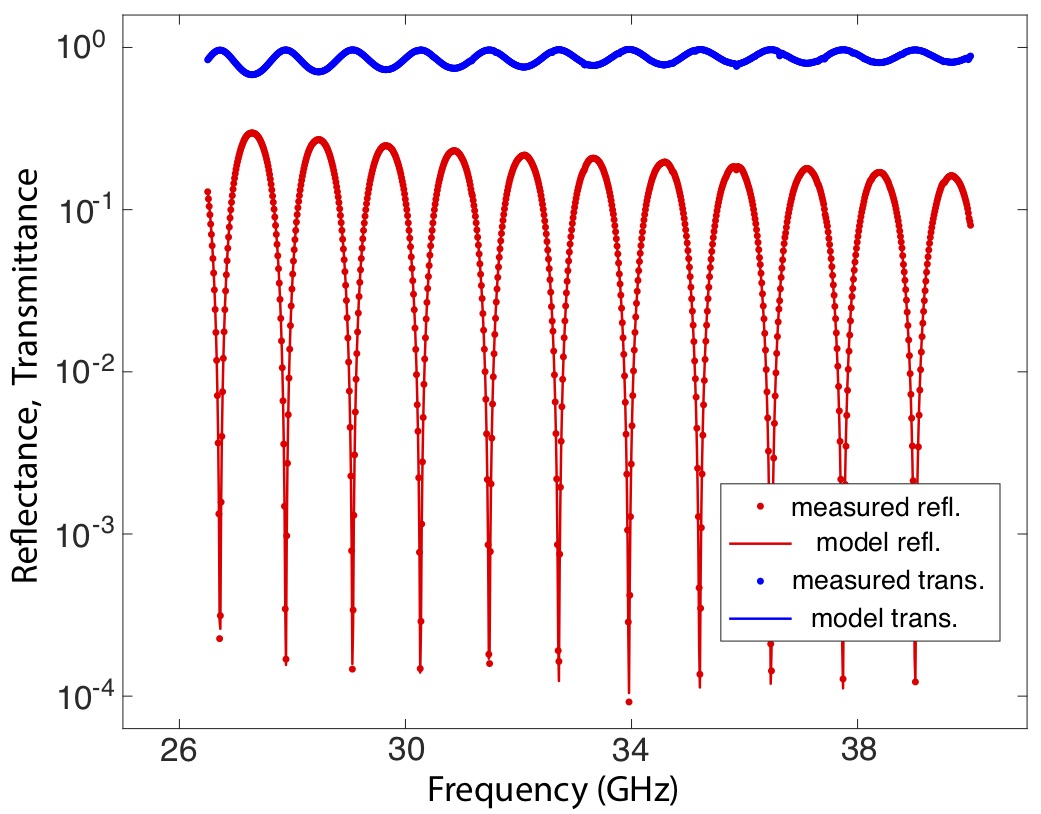} 
\caption{The measured reflectance (red) and transmittance (blue) for a 76.2~mm long WR28 E-plane split-block fixture filled with the stack of 13 etched-PTFE sheets. The solid lines represent the best-fit model.}
\end{figure}

In the measurement setup, the sample-under-test was placed approximately at the location of an output waist of the beam focused by an ellipsoidal mirror. The beam is at its narrowest with a spot size of radius $w_\mathrm{0}$ at the output beam waist. The beam diverges away from the waist approximately following quasi-optical propagation of Gaussian beams~\cite{Goldsmith1998}.  Deviations from the plane-wave approximation near the focus leads to a systematic error in the determination of dielectric parameters~\cite{1017671}, which increases in relative magnitude as $\epsilon^{\prime}~\rightarrow~1$, and with increasing sample thickness and decreasing beam waist radius relative to wavelength, $k_\mathrm{o} w_\mathrm{o}~=~2\pi w_\mathrm{o}/\lambda$. For the measurements presented, $k_\mathrm{o} w_\mathrm{o} \approx $10.4 at 180 GHz, which is large compared to unity. This enables a perturbative correction to the measured reflectance data, $\Gamma_\mathrm{meas}$, following Petersson and Smith ({\it e.g.,} see Eq. 57)~\cite{1017671},
\begin{eqnarray}
\Gamma_\mathrm{corr} \ = \Gamma_\mathrm{meas} \cdot 
\begin{dcases}
\begin{drcases}
\frac{ 1 - \displaystyle\frac{a_\mathrm{1}}{(\displaystyle {k_\mathrm{o} w_\mathrm{o}})^\mathrm{2}} + \displaystyle\frac{a_\mathrm{2}}{(\displaystyle {k_\mathrm{o} w_\mathrm{o}})^\mathrm{4}} }{1 + \displaystyle\frac{1}{(\displaystyle {k_\mathrm{o} w_\mathrm{o}})^\mathrm{4}}}
\end{drcases}\
\end{dcases}.\
\label{Eq:correction}
\end{eqnarray}
The derivation of this expression does not account for the variation in beam waist radius as a function of wavelength at the position of the sample. This effect was accounted for by replacing $k_o w_o$ in the above equation with $k_o w_o \sqrt{1 + (c(z) / k_\mathrm{o} w_\mathrm{o})^2}$, where $c(z) = 2z/w_o$ is a function of the beam waist position, $z$,  relative to the surface of the sample for the optical test configuration~\cite{Goldsmith1998}. Given the measurement's limited spectral range $c(z)$ was treated as a third constant, $a_\mathrm{3}$, in the analysis of the data. 

A least squares fit was performed in log space to accurately identify the locations of the nulls in reflectance. In fitting the data to extract the detailed optical response a simple analytical model was fit at normal incidence with free parameters $\epsilon^{\prime}$, $\epsilon^{\prime\prime}$, and thickness $t$. The thinnest sample is least prone to the systematic bias described above and was used to extract $\epsilon^{\prime}$, which was adopted for the subsequent analysis. The analytical model was then used to fit the measured reflectance of each of the thicker samples after scaling the model by the inverse of the correction function. Physically this corresponds to adopting a common dielectric function for all of the samples characterized. The free parameters while performing these fits were $\epsilon^{\prime\prime}$, thickness $t$, and the numerical coefficients in the correction function. 

In the data reduction procedure described, the observations are effectively mapped from a focused to collimated space, which allows a direct comparison to the modeled plane wave response. In practice, the reflectance response is bounded by focused and collimated cases. In the flight configuration of the PIPER telescope, the vacuum window is neither at a focus nor at a pupil, albeit closer to the collimated case. Table I lists the best-fit model parameters for the bulk samples of four different thicknesses and Fig.~4 shows measurements for two of the samples along with the analytical models.

\begin{figure*}
\centering\includegraphics[width=0.95\linewidth,keepaspectratio]{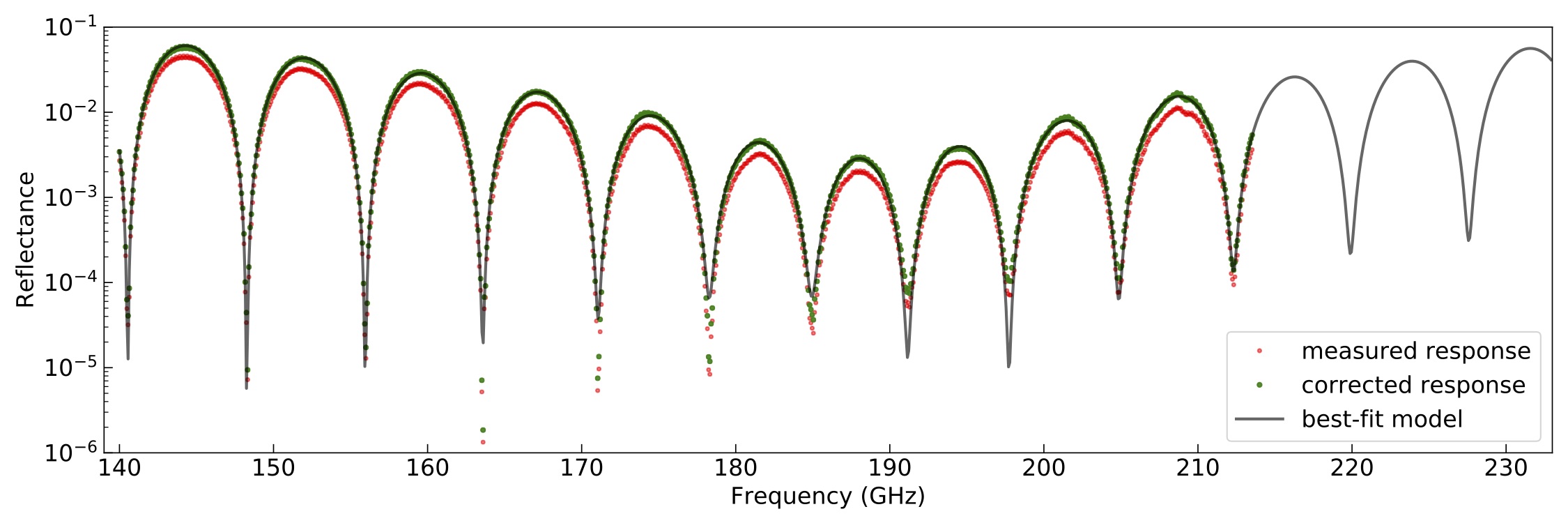}
\caption{ Measured reflectance (red) and corrected reflectance (green) from the AR-coated window with measured total thickness 10.16~mm. The best-fit analytical model is also shown. The model is plotted up to 233 GHz which is the upper half-power band edge of the PIPER 200 GHz observing band. The reflectance spectra as well as the analytical model have identical sampling and are smoothed in the same way as Fig. 4. } 
\end{figure*}

Multiple measurements of the dielectric loss tangent of fused silica over 90 to 250 GHz spanning 0.00028$~<~\tan\delta_\mathrm{e}~<~$0.0026 have been reported~\cite{Goldsmith1998}, with the IR grade fused silica measured in Komiyama {\it et al.}~\cite{88556} being at the lower end of this distribution. The measurements of the loss in the industrial grade fused silica glass presented are consistent with this range.

\subsection{Dielectric Characterization: Bulk PTFE} 
The AR coating material that employed, a PTFE sheet etched on one surface, was also measured. The sheet was measured to have a nominal thickness of 271~$\mu$m. The sample was cut into strips 7.1~mm wide and the resulting sheets were inserted into an E-plane split-block waveguide fixture 76.2~mm in length. After insertion the material was trimmed flush to the waveguide flanges. The final stack consisted of a total of 13 layers with the etched surfaces of the PTFE facing the same direction. The VNA was calibrated with WR28 waveguide TRL (Thru-Reflect-Line) standards and used to measure the response of the Fabry-Perot resonator formed by the impedance contrast between empty guide and the fixture filled with layers of PTFE over a spectral range spanning 26--40\,GHz. 

Following the experimental methods described in Wollack {\it et al.}~\cite{2008IJIMW..29...51W} the structure's frequency response was evaluated as a function of complex dielectric function of the PTFE, $\hat{\epsilon}_{r,\mathrm{PTFE}}$. The minimum deviation between the model and observations indicates $\hat{\epsilon}_{r,\mathrm{PTFE}}$ = 2.01\ +\ 0.0009$i$ ({\it{i.e.}}, tan$\delta_{e} \approx$ 0.0004) in agreement with prior measurements~\cite{Goldsmith1998}. Fig.~5 shows the measured reflectance and transmittance along with the best-fit model. Measurement of a waveguide sample in a fixture 50.8~mm long yielded a consistent result for $\hat{\epsilon}_{r,\mathrm{PTFE}}$. Chemically etching the PTFE increases the material's porosity and lowers the effective index of the layer's surface~\cite{Niklasson:81,Shivola:1999}. If residues of the etchant remain in the pores or delinkage of the polymer structure leads to changes in the material properties, an increase in loss could also occur. 

\begin{table}
\caption{Physical and electrical thicknesses at 180 GHz of the various layers in the dielectric stack. For the window, it is the measured thickness. For the AR and bond layers, the values are of the target thicknesses. }
\begin{center}
\begin{tabular}{lclclclclclcl}
\hline
& & & $t$ (mm) & & & $\epsilon^{\prime}$ & & & $2\pi t\sqrt{\epsilon^{\prime}}/\lambda_\mathrm{0}$ (rad) \\ 
\hline
AR (PTFE) & & & 0.254  & & & 2.02 & & & 1.4 \\
Bond layer & & & 0.013 & & & 3.7 & & & 0.1 \\
Window & & & 9.58 & & & 3.84 & & & 70.2 \\
\hline                 
\end{tabular}\\
\end{center}
\end{table}
\label{tab:model_guess}
\begin{table*}
\caption{Best-fit model parameters for the various layers in the dielectric stack representing the AR-coated window. The imaginary component of relative dielectric function for each layer is held fixed during fitting and values from the measurement of bulk materials were adopted as priors. The uncertainties on $t$ and $\epsilon^{\prime}$ are standard errors on the fit parameters. }
\begin{center}
\begin{tabular}{lclclclcllclcl}
\hline
&&& $t$ (mm) &&& $\epsilon^{\prime}$ &&& $\epsilon^{\prime\prime}\times 10^{3}$ &&& $\tan\delta_{e}$ $\times 10^{4}$ \\ 
\hline
Layer 1 (AR, PTFE) &&& 0.265$\pm$0.002 &&& 2.02$\pm$0.01 &&& 0.9 &&& 4$\pm$0.1 \\
Layer 2 (Bond layer) &&& 0.042$\pm$0.004 &&& 3.27$\pm$0.03 &&& 100 &&& 310$\pm$3 \\
Layer 3 (Window, fused silica) &&& 9.595$\pm$0.03 &&& 3.78$\pm$0.04 &&& 1.7 &&& 4$\pm$0.1 \\
Layer 4 (Bond layer) &&& 0.048$\pm$0.003 &&& 3.27$\pm$0.03 &&& 100 &&& 310$\pm$3 \\
Layer 5 (AR, PTFE) &&& 0.267$\pm$0.002 &&& 2.02$\pm$0.01 &&& 0.9 &&& 4$\pm$0.01 \\
\hline                 
\end{tabular}\\
\end{center}
\end{table*}
\label{tab:model_fit}

\subsection{AR-Coated Window Measurement and Model Fitting}
A prototype silica window AR-coated on both surfaces was characterized using the same VNA test configuration shown in Fig.~3. Using the best-fit complex dielectric function of the bulk silica and the PTFE from above as well as the measured layer thicknesses to set constraints on the model parameters, an analytical model applicable for normal incidence was fit to the measured reflectance data with the various layer thicknesses and real parts of the dielectric function, $\epsilon'$, as free parameters and the imaginary parts, $\epsilon''$, fixed. A multiplicative correction function, see Eq.~\ref{Eq:correction}, was applied to the measured reflectance spectra prior to fitting the data. 

When the electrical delay in the material, $2\pi t\sqrt{\epsilon^{\prime}}/\lambda_\mathrm{0}$, is on the order of a wavelength or smaller, the relative error in the inferred $\epsilon^{\prime}$ is negligible~\cite{1017671}. See Table II for a summary of the layers in the dielectric stack. The aforementioned error should be negligible for the relatively thin AR and glue layers, whereas for the much thicker silica window, the error needs to be corrected. Therefore, the correction function for the 9.69~mm thick bulk sample (i.e, closest to the prototype AR-coated sample) was used to scale the measured reflectance data prior to fitting an analytical model.  

\begin{figure}
\centering\includegraphics[width=0.95\linewidth,keepaspectratio]{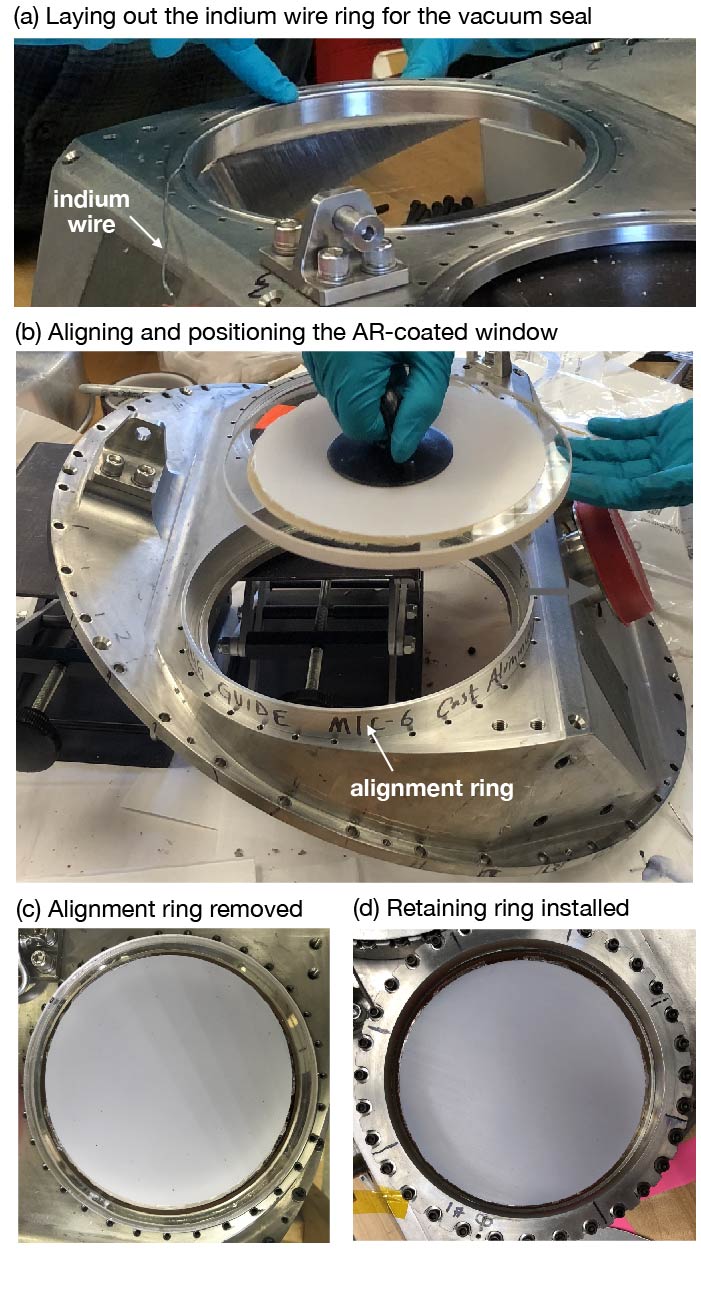}
\caption{Vacuum window installation on the receiver: (a) indium wire ring being laid out in the groove on the window mount frame after cleaning the indium and all surfaces, (b) an alignment ring was positioned and the cleaned AR-coated window was lifted using a suction cup and carefully inserted through the alignment ring to evenly make contact with the indium, (c) alignment ring removed, and finally (d) the retaining ring was aligned and secured using the fastening configuration described in Appendix B.} 
\end{figure}

The AR-coated window was modeled as a five-layer structure where the net reflectance for normal incidence at the $j$-th interface of a stack of multiple dielectric plane parallel layers is given by the recursion relation~\cite{Orfanidis} defined in Eq.~\ref{eq:ref}. In order to reproduce the observed response while minimizing the number of free parameters, the symmetry in dielectric function for layers in the optical stack was enforced, however, it was determined to be necessary to allow differing bond layer and AR layer thicknesses in fitting the data. Fig.~6 shows the measurement along with the best-fit model, and Table III lists the best-fit model parameters.  The data primarily constrains the total loss in the dielectric stack. To address this degeneracy, the imaginary components of relative dielectric functions derived from measurements of the bulk samples were adopted as priors in modeling the data. 

The AR coating bond layer, which is effectively comprised of the Epo-Tek~301-2 epoxy, the Lord AP-134 adhesion promoter and defined by the surface roughness of the etched PTFE, was best fit by $\hat{\epsilon}_{r,\mathrm{bond}} \simeq 3.27$. For comparison, the value reported for a bulk sample of Epo-Tek~301-2 epoxy in Munson {\it et al.}~\cite{Munson:17} is $\simeq 3.7$. While it is challenging to precisely quantify the effective dielectric function in this setting, the inferred value is in reasonable agreement considering the media as a dielectric mixture~\cite{Niklasson:81,Shivola:1999}.

The PIPER 171--233 GHz instrument bandwidth is not fully covered by the 140--214 GHz spectral range of the VNA measurement. The minimum in the envelope of the measured reflectance occurs around 188 GHz. This is shifted to a lower frequency than the targeted $\sim$200~GHz center frequency for the coating and can be attributed to the properties of materials employed and the realized layer thicknesses. However, the fraction of the power reflected averaged over the 171--233 GHz half-power bandwidth as computed from the best-fit model is $<9.6~\times~10^{-3}$ at normal incidence. Using the best-fit model parameters listed in Table III the estimated fraction of incident power absorbed by the AR-coated window at 200~GHz is~$\sim$0.1.

\section{Window Installation on the Receiver and Superfluid-tight Indium Seal}
Vacuum windows appropriate for use with a receiver that is submerged in liquid helium require seals that are impervious to superfluid helium. In this work an O-ring geometry with an indium wire seal was adopted to provide this function. Indium O-ring seals have been widely used in cryogenic systems due to their reliability and ease of construction. Early investigations by Belser~{\it{et~al}}~\cite{Belser:1954} considered the adhesion of indium to a variety of materials including silicon, quartz, and aluminum oxide. Cryogenic vacuum-tight seals for metal-to-glass joints have been developed for a variety of instrumentation applications and reported in the literature~\cite{Willis:1958, Horwitz:1961, Lipsett:1966, Lucas:1959, Abraham:1976, Lim:1986, Haycock:1990}. More recently, D{\"o}ge and Hingerl \cite{Doge:2018} described the use of a hydrogen-tight cryogenic vacuum seal employing indium wire gaskets for amorphous silica windows. A review of design considerations for indium O-ring seals can be found in Turkington and Harris-Lowe~\cite{Turkington:1984}.

The key steps in the window assembly and installation procedure used here are summarized in Fig.~7. A clean pure indium wire 1.27~mm in diameter was arranged in the circular groove as shown in panel (a) of Fig.~7. At the location where the ends of the wire overlap, the wire was cut on a long slant and the cut ends were overlapped and pressed together to form a firm ring. An alignment ring was positioned on the seal area. The window was lifted using a suction cup as shown in panel (b) of Fig.~7 and inserted through the alignment ring, making contact with the indium as evenly as practicable. 

The alignment ring was then carefully removed taking special care not to bump the window out of its centered position on the indium wire ring and window aperture. Fig.~7 panel (c) shows the window in place and the alignment ring removed. Window mount frames can be designed to compensate for the differential coefficient of thermal expansion (CTE) of the windows and frame~\cite{Miura:1982} both radially and in thickness. However, in this case, the required compensation washer results in a large mount frame that interferes with the beam. 

Therefore, the design described achieves compliance in the frame using four stacked belleville washers in series. The bellevilles both provide sufficient force to compress the indium seal and retain necessary spring travel to take up differential CTE in cryogenic operation without losing compliance and mechanically constraining the frame against the window. The interface between the window and frame has an aluminum washer to prevent force concentrations at any roughness or contamination at the mount interface. Fig.~7 panel (d) shows the retaining ring secured with fasteners to complete the window installation. A complete step-by-step window assembly sequence can be found in Appendix B.

\begin{figure}
\centering\includegraphics[width=0.95\linewidth,keepaspectratio]{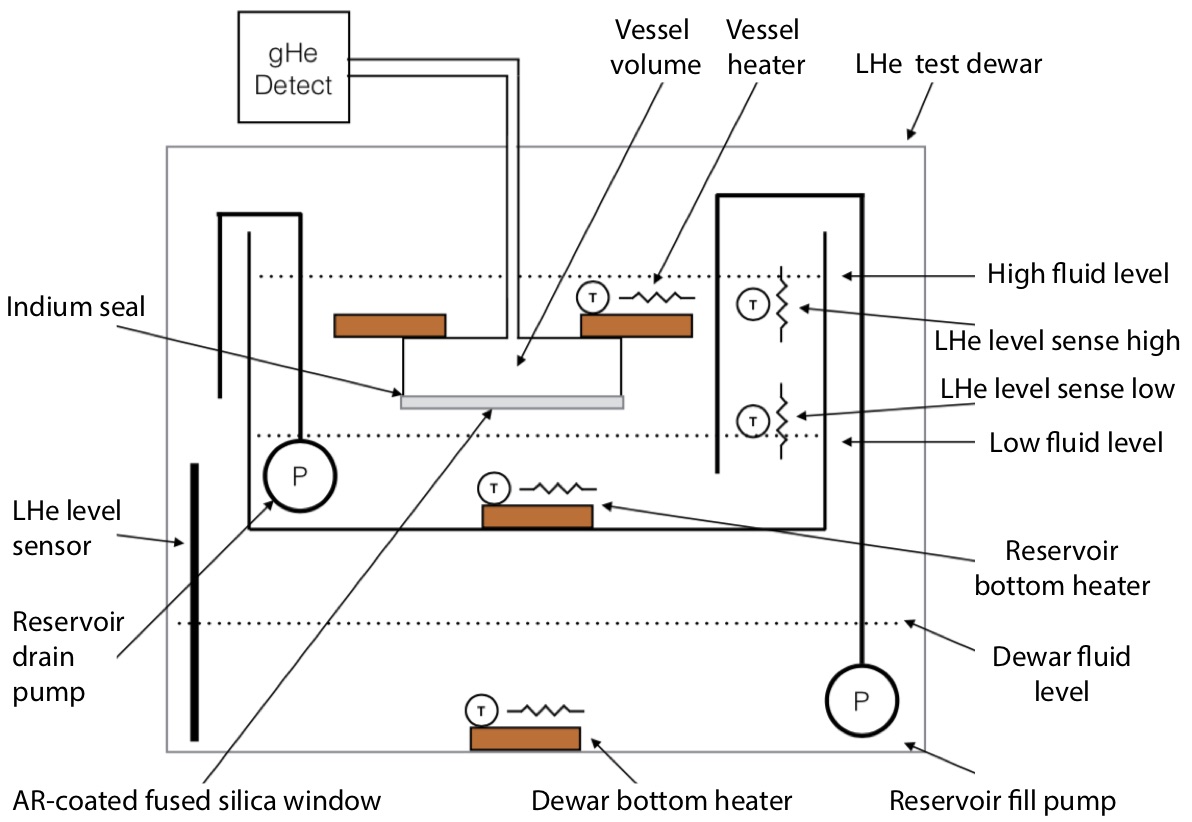}
\caption{Setup for testing the superfluid helium tight window seal. The window was installed on a test vessel. The test vessel was mounted in a liquid helium (LHe) test dewar and the test vessel volume wass pumped out to create a vacuum. A leak detector was then connected to the test vessel.}
\end{figure}

\section{Window Seal Leak Testing and Thermal Cycling}
\vskip1sp
The window was installed on a test vessel using a indium wire ring seal described in Section~V. The test vessel was mounted in a liquid helium (LHe) test dewar, surrounded by a reservoir inside the dewar. The test vessel was pumped out to create a vacuum and subsequently connected to a leak detector. Fig.~8 shows a schematic of the leak-test setup. The window was then cycled 50 times from ambient to liquid nitrogen (LN2) temperatures without window or seal failure, as confirmed by helium gas leak detection rates undetectable above the measurement floor of 10$^{-8}$ mbar L/s at ambient temperature after cycling. 

Next, the vacuum seal to superfluid helium was tested. The test procedure went as follows. The test dewar was partially filled with LHe and pumped out to simulate the ambient pressure at balloon altitudes. The superfluid helium was then pumped into the upper reservoir using a superfluid fountain-effect pump until its level rises above the window test unit. After waiting for several minutes ({\it i.e.,} to enable the possibility of superfluid helium leakage into the test vessel), the superfluid was pumped back out of the upper reservoir. 

The window test unit was subsequently heated to mobilize any helium inside the test vessel. Then the heat was turned off, the leak rate was noted, and the above steps were repeated. Helium liberated in the heat mobilization was consistent with residual adsorbed gas in the test device, and diminished with each test cycle. The observed leak rates gives a time scale on the order of several hundreds of days for the receiver to reach up to a pressure at which it would no longer function.

\section{Summary}
\vskip1sp
An anti-reflection coated fused silica vacuum window has been demonstrated for the PIPER balloon-borne instrument. Laboratory measurement of the AR-coated window was consistent with the analytical model presented and confirmed the reduction in net reflectance at the window to below a percent averaged over the 171--233\,GHz half-power bandwidth of the instrument observing in the 200 GHz band. The window survived multiple thermal cycling tests down to liquid nitrogen temperatures. Leak rate measurements met the operational requirements. A quasi-optical measurement technique in a focused beam was developed and used to validate the AR-coated window optical performance. A vacuum seal that is robust to superfluid helium leaks was realized using indium wire ring. The window assemblies were installed on the receiver and operated successfully during PIPER's flight~\cite{Essinger2020Primordial} from New Mexico in October, 2019. 

\begin{acknowledgments}
R. Datta$^{\prime}$s research was supported by an appointment to the NASA Postdoctoral Program at the NASA Goddard Space Flight Center (GSFC), administered by Universities Space Research Association under contract with NASA. K. Helson’s research was supported by NASA under award number 80GSFC17M0002. PIPER is supported by the grant 13-APRA13-0093. Support for development of the microwave instrumentation and metrology was provided by the GSFC Internal Research and Development (IRAD) program. The authors acknowledge the collective contributions of the anonymous reviewers which greatly improved the clarity of the final presentation. This article may be downloaded for personal use only. Any other use requires prior permission of the author and AIP Publishing. This article appeared in Rev. Sci. Inst. Volume 92, Issue 3 and may be found at https://doi.org/10.1063/5.0029430.
\end{acknowledgments}

\begin{appendix}
\section{AR Coating Procedure}
\begin{figure}
\centering\includegraphics[width=0.95\linewidth,keepaspectratio]{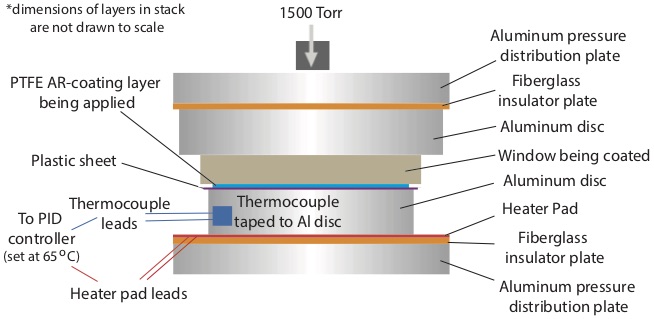}
\caption{Schematic of the stack of materials in the hydraulic press used for applying the PTFE AR coating layer on the window. Only one surface of the window ({\it{i.e.,}} the bottom surface) was coated at a time. The thickness of the thin layers in the stack are exaggerated for clarity.}
\end{figure}

The procedure adopted for applying the AR coating on the fused silica windows is described here. The AR coating layer was glued to the window under pressure using a hydraulic press in two steps for the two surfaces. Fig.~9 shows a schematic of the various components in the stack. After wiping down all surfaces of the window with acetone and isopropanol, it was placed on a $\sim$23~cm diameter aluminum disc with the surface to be coated facing up. A Kapton\footnote{www.dupont.com/electronic-materials/kapton-polyimide-film.html} tape sheet was then applied covering the entire window surface. Next, an aluminum ring (outer diameter = 19~mm, inner diameter = 16.5~cm) was placed on top such that the outer edge of the ring lined up with the edge of the window. Then, the Kapton was cut out along the inner diameter of the aluminum ring with a razor blade and peeled off. The aluminum ring was removed and the exposed surface of the window was again cleaned with acetone and isopropanol. The Kapton tape sheet was to protect the edges of the window from excess epoxy flowing into the seal area.  

A 16~cm diameter sheet of the 254~$\mu$m thick etched PTFE was cut out and both surfaces were cleaned with acetone and isopropanol. The PTFE was placed on another aluminum disc of diameter $\sim$17~cm with the etched surface facing up. Next, the Lord AP-134 adhesion promoter was applied to both the window surface to be coated and the etched surface of the PTFE separately using a syringe to dispense the Lord AP-134. A foam-tipped applicator was used to quickly spread it around the surface leaving only a thin (1-5-2.5 $\mu$m) layer. This step required care as the AP-134 starts to cure and become tacky relatively quickly. These items were left undisturbed for an hour. The Epo-Tek~301-2 epoxy was then mixed in the recommended proportions of 3.0 g of Part A per 1.05 g of Part B and outgassed under vacuum for an hour. Approximately 0.75~mL of the epoxy was spread on the window surface in a spiral pattern starting at the center using a syringe. This amounts to approximately 4~$\mu$L/cm$^{2}$. The PTFE sheet was placed and centered on top of the prepared window surface with the etched side facing down. A plastic (LDPE) sheet was placed on top to protect from epoxy spill outside the PTFE edges. The aluminum disc $\sim$17~cm in diameter was then placed on top while centering it by eye. 

An aluminum pressure distribution plate was placed on the foot of the press and a ceramic insulator tile was placed on the plate. A heater pad from McMaster Carr was placed on top of the tile and used to decrease the curing time of the epoxy. The entire prepared stack from the previous paragraph was then inverted and placed on top of the heater pad and another ceramic insulator tile was placed on top of the stack followed by a second aluminum pressure distribution plate. A thermocouple was taped to the side of the $\sim$17~cm diameter aluminum disc. A pressure of 1500~Torr was applied from the top with the hydraulic press and the stack was left to cure under heat for at least 12 hours. The temperature was controlled using a PID controller set to 65$^{0}$C and a variable AC transformer (Variac\footnote{www.variac.com}). Once done, the stack was disassembled and inspected, and the coating was repeated on the other side. 

\begin{figure*}
\centering\includegraphics[width=1\linewidth,keepaspectratio]{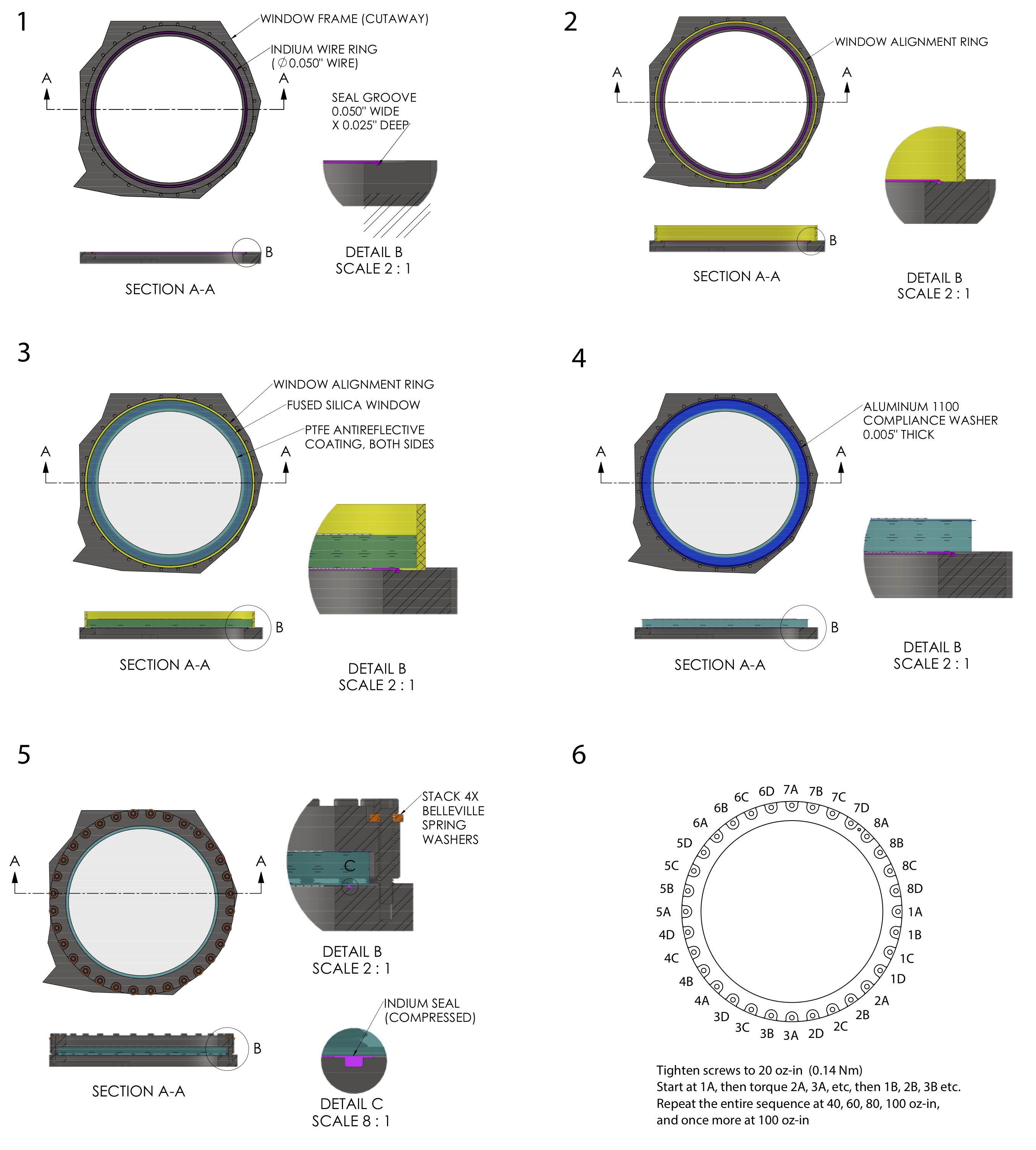}
\caption{Window assembly sequence. Schematics of a cutaway of the vacuum window frame along with the indium wire ring, the window alignment ring, the aluminum compliance washer, and the PTFE-coated window are labelled. The numbering of the panels are indicative of the stepwise window installation procedure described in the text.}
\end{figure*}

\section{Window Installation Procedure}
The sequence and techniques used for installing the AR-coated window on the PIPER receiver are described here. When installing for the first time on the machined stainless steel receiver structure, the parts were first degreased by wiping down with 1:1 Crystal Simple Green and deionized water, then with acetone, and then with isopropanol. Both the exposed silica surface of the window and the stainless steel in the area of the indium groove were cleaned using wipes and cotton swabs, paying particular attention to the indium groove. 

Pure indium wire 1.27~mm in diameter was cleaned with wipes and isopropanol and arranged in the groove. At the location where the ends of the wire overlap, the wire was cut on a long slant using a razor blade that has been wiped down with isopropanol. The cut ends were overlapped and pressed together using firm finger pressure to weld the ends, with the weld positioned high on lid face at 12 o$^{\prime}$clock. An alignment ring was positioned on the seal area. The window was lifted using a suction cup and inserted through the alignment ring, making contact with the indium as evenly as practicable as shown in step 3 of Fig.~10. Then, the alignment ring was carefully removed taking special care not to bump the window out of its centered position on the indium wire ring and window aperture. 

A thin (127~$\mu$m) foil (aluminum 1100) compliance washer was placed between the glass and a retaining ring. The foil surface-compliance washer was put in to take up any variation in the inside surface of the retaining ring left over from the machining process. This provides protection against dust particulates from reaching the region between window and the retaining ring during assembly. The retaining ring was mounted to a tapped thermometer mounting hole. Compliance in the frame was achieved with a stack of four Number 8 belleville disk springs\footnote{www.mcmaster.com/catalog/125/1336} (302 stainless, 0.190$^{\prime\prime}$/0.375$^{\prime\prime}$ ID/OD, 0.020$^{\prime\prime}$ thick) in series on each no. 8-32$^{\prime\prime}$ black oxide coated stainless steel socket cap head screw. The pre-load as a function of displacement on the belleville washers was determined using an annular load cell from Omega\footnote{www.omega.com}, the LC8200-500-3K, and a dial indicator. A single Belleville washer\footnote{www.mcmaster.com/metal-belleville-spring-washers} was observed to be entirely flattened by a $65$\,N load. A stack of three washers, all with the cone pointed towards the head of the screw and the same pre-load, had $36$\,$\mu$m of remaining travel. This was sufficient for thermal compliance, but to provide increased margin, a fourth washer in the same orientation was added to the stack. A small amount of NS-10 anti-seize lubricant from Bostik, Inc.\footnote{www.bostik.com} was applied to the threads of the screws, both during testing and during window installation on the flight receiver, to try to get consistent torque to pre-load behavior. One drop was applied near the threaded end, smeared around the circumference. New washers were used for each installation.  

A fastening configuration was required that would provide consistent loading of the indium seal in the $2$\,K to $300$\,K temperature range. Lack of compliance in the fastening led to window failure at warm up during initial testing. Sufficient pre-load to compress the indium seal was also required and it was desirable to leave the fasteners in high pre-load after compression.

The fastener installation procedure adopted for the window assembly is described next. The distance from the threaded surface to the head of the screw was 19.5~mm when the washers were uncompressed. That left 5.9~mm length of engagement, which was about seven threads. Each screw was lightly finger tightened, alternating between screws across the window diameter to keep the loading even. An over-torque-slip type torque screwdriver was used to apply torque to groups of four screws as illustrated in step 6 of Fig.~10. The level of torque was incrementally raised as follows: 0.14, 0.28, 0.42, 0.56, 0.70, 0.70 (repeated) N-m. During visual inspection, the criteria for good seal formation was the observation of shiny indium extrusions, clearly wet to the surface of the window, to greater than three times the diameter of the indium wire. A compression force of 145 lbs (650 N) at each screw, using the careful tightening sequence as described above, was appropriate. For the 32 screws, that created a total load of 4600 lbs (20,700 N). The pressure on the final seal was 1400 PSI (72,400 Torr).

\end{appendix}

\vskip 10mm

\textbf{DATA AVAILABILITY STATEMENT}
\vskip 5mm
\noindent
The data that support the findings of this study are available from the corresponding author upon reasonable request.

\vskip 10mm

\cite{*}

\bibliography{main_manuscript}

\end{document}